\documentclass[12pt]{article}
\usepackage{amsmath, amsthm, amsfonts}

\newif\ifpdf
	\ifx\pdfoutput\undefined
	\pdffalse 
\else
	\pdfoutput=1 
	\pdftrue
\fi
\renewcommand{\labelenumi}{(\arabic{enumi})}   

\newcommand{\Eq}[1]      {(\ref{#1})}

\newcommand{\Th}[1]      {Thm.~\ref{#1}}
\newcommand{\Lem}[1]    {Lem.~\ref{#1}}

\newcommand{\Sec}[1]    {\S\ref{#1}}
\newcommand{\Tbl}[1]    {Table~\ref{#1}}
\newcommand{\Prop}[1]  {Prop.~\ref{#1}}

\newcommand{\rev}[1]    {({\textbf{R\ref{#1}}})}
\newcommand{\Naa} {\textbf{(AA)}}
\newcommand{\Nea} {\textbf{(EA)}}
\newcommand{\Nee} {\textbf{(EE)}}
\newcommand{\G}{{\mathcal G}}
\newcommand{\A}{{\mathcal A}}
\newcommand{\calS}{{\mathcal S}}
\newcommand{\E}{{\mathcal E}}
\newcommand{\R}{{\mathbb R}}

\newcommand{\C}{{\mathbb C}}

\newcommand{\Henon}{H\'enon }

\newtheorem{thm}{Theorem}
\newtheorem{lem}[thm]{Lemma}

\newtheorem{prop}[thm]{Proposition}

\theoremstyle{definition}

\newtheorem*{rem}{Remark}

\begin{document}

\title{Reversible Polynomial Automorphisms of the Plane: the Involutory Case}
\author{A. G{\'o}mez\thanks{
       Support from the CCHE Excellence grant for Applied
       Mathematics  is gratefully acknowledged.}\\
       {\small \begin{tabular}{c}
       Department of Mathematics\\
       University of Colorado\\
       Boulder CO 80309-0395\\
       and\\
       Departamento de Matem\'aticas\\
       Universidad del Valle, Cali Colombia
       \end{tabular}} 
       \and J. D. Meiss\thanks{ Support from NSF grant DMS-0202032 is
       gratefully acknowledged.  JDM would also like to thank H.
       Dullin and J. Roberts for helpful conversations.}\\
       {\small \begin{tabular}{c}
       Department of Applied Mathematics\\
       University of Colorado\\
       Boulder CO 80309-0526
       \end{tabular}}}
\date{\today}
\maketitle

\begin{abstract}
     Planar polynomial automorphisms are polynomial maps of the plane
     whose inverse is also a polynomial map.  A map is reversible if
     it is conjugate to its inverse.  Here we obtain a normal form for
     automorphisms that are reversible by an involution that is also
     in the group of polynomial automorphisms.  This form is a
     composition of a sequence of generalized \Henon maps together
     with two simple involutions.  We show that the coefficients in
     the normal form are unique up to finitely many choices.
     \end{abstract}

\vspace*{1ex}
\noindent

\section{Introduction}
A diffeomorphism $g$ has a reversing symmetry, or is ``reversible'' if
it is conjugate to its inverse
\cite{deVolgeleare58,Devaney76,Sevryuk86, Lamb98}, i.e., there exists
a diffeomorphism $R$ such that
\begin{equation}\label{eq:reversible}
       g^{-1} = R\,g\,R^{-1} \;.
\end{equation}
Reversible maps have a number of special properties and occur often in
applications.  For example, if the phase space of a system consists of 
configuration coordinates and momenta, then it is often the case that 
reversal of the momenta, $R(q,p) = (q,-p)$, corresponds to the reversal 
of time.  Note that if $R$ is a reversor, then it generates a family of 
reversors $g^{n}\,R$, for any integer $n$.

Reversibility is often associated with Hamiltonian systems, and more
generally to conservative dynamics.  Most representative examples of
reversible systems originate in the study of Hamiltonian dynamics, and
many Hamiltonian systems appearing in applications are reversible. It
is, however, well known that these properties are independent:
Hamiltonian systems need not be reversible, and reversible systems
need not be conservative, or even volume-preserving \cite{MacKay93} 
(see also \cite {Lamb98} and references therein).

Often the reversor is an involution, $R^{-1} =R$. This is the
case, for example, for the physical reversor mentioned above. When this is
true, a reversible map can be written as the composition of two
involutions
\[
       g = (gR)\,(R) \;,
\]
since $gR$ is also an involution.  For example, the standard
\cite{Greene79} and area-preserving \Henon maps \cite{Devaney76} have
involutions as reversors.

Orbits of $g$ that intersect the fixed sets of the family of reversors,
$g^{n}R$, are called ``symmetric orbits.'' Since $\mbox{Fix}(R) = \{z:
R(z) = z\}$, generally has lower dimension than the phase space, these
orbits are special, and are easier to find than general orbits. Their
bifurcations are also special; for example in the two-dimensional
case, pitchfork bifurcations are generic in one-parameter families of
reversible maps \cite{Rimmer78}. From this point of view, a reversor 
that is an involution is particularly interesting, because its fixed set is
often nontrivial. For example, every orientation-reversing involution of the 
plane has a one-dimensional fixed set \cite{MacKay93}.

Our goal is to classify reversible polynomial automorphisms of the
plane.  Polynomial maps form one of the simplest, nontrivial classes
of nonlinear dynamical systems.  A polynomial automorphism is a
polynomial diffeomorphism whose inverse is also polynomial.  Such maps
can give rise to quite complicated dynamics as exemplified by the
renowned \Henon quadratic map \cite{Henon69}. A larger family of maps
consists of the generalized \Henon maps of the form
\begin{equation}\label{eq:Henon}
	h:(x,y)\rightarrow (y,p(y)-\delta\,x),
\end{equation}
where $\delta  = \det(Dh) \neq 0$ is the Jacobian of the map and 
$p(y)$ is any polynomial of degree $\ge 2$.  These maps are reversible 
when they are area and orientation-preserving, i.e., $\delta = 1$, or
area-preserving and orientation-reversing, $\delta =-1$, providing
that $p$ is an even polynomial.  Some of their dynamics has been studied in
\cite{Friedland89, Dullin00}.

A remarkable property of any polynomial automorphism is that the
determinant of its Jacobian matrix is a nonzero constant (a famous
unsolved question---the Jacobian conjecture---is to determine if all
polynomial maps with a nonzero, constant Jacobian are polynomial
automorphisms \cite{Essen95}). It follows as a simple consequence of
\Eq{eq:reversible} that when the Jacobians of $g$ and $R$
are constant, then $g$ is volume-preserving. Therefore polynomial
automorphisms possessing polynomial reversors must be
volume-preserving.

Since the composition of any two polynomial automorphisms is is again
a polynomial automorphism, the set of polynomial automorphisms of
$\R^{2}$ or of the complex plane, $\C^{2}$, is a group; we call it
$\G$.  We will make use of the results of Jung on the structure of
$\G$ to investigate polynomial automorphisms that are reversible.  Our
results also heavily use the normal form for elements of $\G$ in terms
of generalized \Henon maps obtained by Friedland and Milnor
\cite{Friedland89}.

For the purposes of this letter, we restrict attention to the case
that $R$ is an involution.  While this is a restriction, we will show
in a forthcoming paper that noninvolutory reversors are exceptional in
the sense that additional conditions on the generalized \Henon
transformations are required \cite{Gomez02}.  It has been previously
shown that all reversible polynomial mappings in generalized standard
form possess involutory reversors \cite{Roberts01}.

We will classify all maps that are reversible by an
involution in $\G$ and obtain unique normal forms for these maps
under conjugacy in $\G$.  Just as in \cite{Friedland89}, these normal
forms involve compositions of generalized \Henon maps; however, in
this case, the reversors are introduced by including two involutions
in this composition:
\begin{equation}\label{eq;genNormalForm}
       (h_{1}^{-1}\ldots h_{m}^{-1})\,r_{1}\,(h_{m}\ldots h_{1})\,r_{0} \;,
\end{equation}
where $r_{0}$ and $r_{1}$ involutions.  These basic involutions are
constructed from ``elementary'' involutions (see \Prop{prop:elem}) or
the simple affine permutation $t(x,y) \rightarrow (y,x)$.  This gives
three cases that we call \Nee, \Nea, and \Naa, see \Th{thm:normal}. 
Conversely, any map of the form \Eq{eq;genNormalForm} is reversible
when the $r_{i}$ are involutions.
\section{Background}
For future reference we include some basic terminology and results 
(see e.g.~\cite{Friedland89} for more details).

We will study polynomial maps of the complex plane, $\C^{2}$, occasionally 
specializing to the real case. We let $\G$ denote the group of  {\em polynomial
automorphisms} of the complex plane i.e.\ the set of bijective  maps
\[
	g:(x,y)\rightarrow\left(X(x,y),Y(x,y)\right),\ \ X,Y \in \C[x,y] \;,
\] 
having a polynomial inverse.  Here $\C[x,y]$ is the ring of
polynomials in the variables $x$ and $y$, with coefficients in $\C$. 
The {\em degree} of $g$ is defined as the largest of the degrees of
$X$ and $Y$.

The subgroup $\E\subset \G$ of {\em elementary maps} consists of maps of the form 
\begin{equation}\label{eq:elem}
	e:\left(x,y\right)\rightarrow \left(\alpha\,x+p(y), \beta\,y +\eta \right) \;,
\end{equation}
where $\alpha \beta \neq 0$ and $p(y)$ is any polynomial.  The
subgroup of affine automorphisms is denoted by $\A$.  The affine maps
that are also elementary will be denoted by $\calS=\A\cap \E$, while
$\hat{\calS}$ will denote the group of diagonal affine automorphisms,
\begin{equation}\label{eq:diagonal}
	\hat{s}:(x,y)\rightarrow \left(\alpha\,x +\xi, \beta\,y +\eta\right) \;.
\end{equation} 

It is worthwhile to note that $\hat{\calS}$ is the largest subgroup of
$\calS$ normalized by $t:(x,y)\rightarrow (y,x)$, i.e., such that
$t\,\hat{\calS}\,t=\hat{\calS}$.  On the other hand a map $s\in \calS$
commutes with $t$ if it is a diagonal automorphism \Eq{eq:diagonal}
with $\alpha=\beta$ and $\xi=\eta$.  This subgroup of $\hat{\calS}$ is
the centralizer of $t$ in $\calS$,
\[
	C_{\calS}(t)=\{s\in \calS:\,s\,t\,s^{-1}=t\} \;.
\]
Finally, conjugacy by $t$ will be denoted by $\phi$,
\begin{equation}\label{eq:conjugacy}
	\phi(g)= t\,g\,t \;.
\end{equation}
Thus if $s\in C_{\calS}(t)$, then $\phi(s)=s$.

According to Jung's Theorem \cite{Jung42} every polynomial automorphism 
$g\notin \calS$, can be written as
\begin{equation}\label{eq:Jung}
	g=g_m\, g_{m-1}\, \cdots \, g_2\, g_1,\quad\quad 
	g_i\in \left(\E\cup \A\right) \setminus \calS,\ i=1,\ldots m \;, 
\end{equation}
with consecutive terms belonging to different subgroups $\A$ or $\E$. An
expression of the form \Eq{eq:Jung} is called a {\em reduced word} of length 
$m$. An important property of a map written in this form is that its degree 
is the product of the degrees of the terms in the composition
\cite[Thm. 2.1]{Friedland89}. As a consequence of this fact it can be seen
that the identity cannot be expressed as a reduced word
\cite[Cor 2.1]{Friedland89}. This fact in turn means that $\G$ is the free
product of $\E$ and $\A$ amalgamated along $\calS$. The structure of $\G$ as an
amalgamated free product determines the way in which reduced words
corresponding to the same polynomial automorphism are related. 

\begin{thm}\label{thm:uniqueness}    
    (cf.  \cite[Cor 2.3]{Friedland89} or \cite[Thm 4.4]{Magnus66}). 
    Two reduced words $g_{m}\,\cdots\,g_{1}$ and
    $\tilde{g}_{n}\,\cdots\,\tilde{g}_{1}$ represent the same
    polynomial automorphism $g$ if and only if $n=m$ and there exist
    maps $s_{i}\in \calS,\ i=0,\dots ,m$ such that $s_{0}=s_{m}=id$
    and $\tilde{g_{i}}=s_{i}\,g_{i}\,s_{i-1}^{-1}$.
\end{thm}

>From this theorem it follows that the length of a reduced word \Eq{eq:Jung} as
well as the degrees of its terms are uniquely determined by $g$. The sequence
of degrees $(l_{1},\dots,l_{n})$ corresponding to the maps $(g_{1},\dots
,g_{m})$, after eliminating the $1$'s coming from affine terms, is called 
the {\em polydegree} of $g$.

A map is said to be {\em cyclically-reduced} in 
the trivial case that it belongs to $\A\cup \E$ or in the case that it can be
written as a reduced word \Eq{eq:Jung} with the additional conditions $m\geq 2$
and $g_m$, $g_1$ not in the same subgroup $\E$ or $\A$.

Two maps $g,\tilde{g}\in \G$ are conjugate in $\G$ if there exists $f\in \G$
such that $g=f\,\tilde{g}f^{-1}$. If $f$ belongs to some subgroup
$\mathcal{F}$ of $\G$ we say that $g$ and $\tilde{g}$ are
$\mathcal{F}$-conjugate. It can be seen that every $g\in \G$ is conjugate to a
cyclically-reduced map. It can also be proved that every affine map $a$ can be
written as $a=s\,t\,\tilde{s}$, with $t:(x,y)\rightarrow (y,x)$ and
$s,\tilde{s}$ affine elementary maps. From these facts it follows that every
polynomial automorphism that is not conjugate to an elementary or an affine
map is conjugate to a reduced word of the form, 
\begin{equation}\label{eq:te-form} 
	g=t\,e_m\,\cdots \,t\,e_{2}\,t\,e_1,\quad e_i\in \E\setminus \calS,\quad 
	i=1,\cdots m,\quad m\geq 1 \;. 
\end{equation}
Moreover this representative of the conjugacy class is unique
up to modifications of the maps $e_i$ by diagonal affine automorphisms and
cyclic reordering. More precisely we have the following theorem 
(cf. \cite[Thm. 4.6]{Magnus66}).

\begin{thm}\label{thm:conjugate}
	Two nontrivial, cyclically-reduced words $g=g_{m}\,\cdots\,g_{1}$ and 
	$\tilde{g}=\tilde{g}_{n}\,\cdots\,\tilde{g}_{1}$ are conjugate if and only if
	they have the same length and there exist automorphisms 
	$s_{i}\in \calS,\ i=0,\dots,m$, with $s_{m}=s_{0},$ and a cyclic permutation,
	\[
		\left(\hat{g}_{m},\ldots ,\hat{g}_{1}\right)=
		 \left(\tilde{g}_{k},\ldots ,\tilde{g}_{1},\tilde{g}_{m}
		 \ldots ,\tilde{g}_{k+1}\right)
	\] 
	such that $\hat{g}_{i}=s_{i}\,g_{i}\,s_{i-1}^{-1}$. In that case, 
	\[
		s_{0}\,g\,s_{0}^{-1}=\hat{g}_{m}\,\cdots\, \hat{g}_{1} \;.
	\]
	
	In particular, if $g=t\,e_m\,\cdots\, t\,e_1$ and $\tilde{g} =
	t\,\tilde{e}_{m}\,\cdots\, t\,\tilde{e}_{1}$ are conjugate,
	there exist diagonal automorphisms $s_{i}\in\hat{\calS},\
	s_{m}=s_{0}$, and a cyclic reordering,
	\[
		\left(\hat{e}_{m},\ldots ,\hat{e}_{1}\right)=
		 \left(\tilde{e}_{k},\ldots ,\tilde{e}_{1},\tilde{e}_{m}
		 \ldots ,\tilde{e}_{k+1}\right),
	\] 
	such that $t\,\hat{e}_{i}=s_{i}\,t\,e_{i}\,s_{i-1}^{-1}$ and 
	\[
		s_{0}\,g\,s_{0}^{-1}=t\,\hat{e}_{m}\,\cdots\, t\,\hat{e}_{1}.
	\]
\end{thm}
\begin{proof} 
Let us consider $g=g_{m}\,\dots\,g_{1}$ and
$\tilde{g}=\tilde{g}_{n}\,\dots\,\tilde{g}_{1}$, two nontrivial,
cyclically-reduced, conjugate words.  By assumption, there is a
reduced word $f=f_{k}\,\cdots\,f_{1}\in \G$, such that
$g=f\,\tilde{g}\,f^{-1}$.  Then,
\begin{equation}\label{eq:conjugate}
	g_{m}\,\dots\,g_{1}=
	f_{k}\,\cdots\,f_{1}\,\tilde{g}_{n}\,\cdots\,\tilde{g}_{1}\,
	f_{1}^{-1}\,\cdots\,f_{k}^{-1}.
\end{equation}
However, the word on the right hand side of \Eq{eq:conjugate} is not reduced. 
Since $\tilde{g}$ is cyclically-reduced, we can suppose, with no loss of
generality, that $f_{1}$ and $\tilde{g}_{n}$ belong to the same subgroup $\A$ or 
$E$, so that $f_{1}^{-1}$ and $\tilde{g}_{1}$ lie in different subgroups. 
Taking into account \Th{thm:uniqueness} and that \Eq{eq:conjugate} represents a 
cyclically-reduced map, we can reduce to obtain
\begin{equation}\label{eq:reduced}
	f_{k}\,\cdots\,f_{1}\,\tilde{g}_{n}\,\cdots\,\tilde{g}_{1}=
	\begin{cases}
		s_{k}\,\tilde{g}_{n-k}\,\cdots\,\tilde{g}_{1} &\text{if }n\geq k\\
		f_{k}\,\cdots\,f_{n+1}\,s_{n} &\text{if } n< k,
	\end{cases}
\end{equation}
where $s_{n}, s_{k}\in \calS$. Moreover there exist $s_{i}\in \calS,\ s_{0}=id$, 
such that $f_{i}\,s_{i-1}\,\tilde{g}_{n-i+1}=s_{i}$ for $i=1,\dots ,\min(n,k)$. 

For the case $n\geq k$, 
\begin{align*}
	g_{m}\,\dots\,g_{1}&=
	s_{k}\,\tilde{g}_{n-k}\,\cdots\,\tilde{g}_{1}\,f_{1}^{-1}\,\cdots\,f_{k}^{-1}\\
	&=(s_{k}\,\tilde{g}_{n-k})\,\tilde{g}_{n-k-1}\,\cdots\,\tilde{g}_{1}\,
	\tilde{g}_{n}\,\dots\,\tilde{g}_{n-k+2}\,(\tilde{g}_{n-k+1}\,s_{k}^{-1}) \;,
\end{align*}
and applying \Th{thm:uniqueness} we have the result. The case
$n<k$ follows analogously. 

To prove the second statement of this theorem it is enough to recall that
given $s\in \calS$, $t\,s\,t$ stays in $\calS$ if and only if $s$ is diagonal.
\end{proof}

It can be noted from the previous theorem that the length of a 
cyclically-reduced word is an invariant of the conjugacy class. Since a 
nontrivial, cyclically-reduced word has the same number of elementary and 
affine terms, we refer to this number as the {\em semilength} of the word. 
\Th{thm:conjugate} also implies that two cyclically-reduced maps that are 
conjugate have the same polydegree up to cyclic permutations. We will call 
this sequence the polydegree of the conjugacy class.

\section{Involutory Reversing Symmetries}\label{sec:involutory}
As was noted in the introduction, a map has an involutory reversing
symmetry if and only if it can be expressed as the composition of two
involutions.  In this section we make use of this property to describe
the class of polynomial automorphisms that possess involutory
reversors.  We start by studying involutions.

\subsection{Polynomial Involutions}
To begin, we show that all polynomial involutions are dynamically 
trivial, i.e., are conjugate to affine or elementary maps.

\begin{prop}\label{prop:inv}
	A map $g\in \G$ is an involution if and only if $g$ is conjugate to an affine 
	or to an elementary involution.
\end{prop}

\begin{proof} 
Assume that $g$ is an involution conjugate in $\G$ to a cyclically-reduced map with
semilength $m:$ $\tilde{g} = a_{m}\,e_{m}\,\cdots \,a_{1}\,e_{1}$, $m\geq 1$. 
As the involution condition is preserved under conjugacy we have,
\[
	\tilde{g}^{2}= a_{m}\,e_{m}\,\cdots
	\,a_{1}\,e_{1}\,a_{m}\,e_{m}\,\cdots\,a_{1}\,e_{1}=id \;. 
\]
But this is a contradiction since the identity cannot be written as a reduced
word. It follows that $g$ must be conjugate to either an affine or to an
elementary map. 
\end{proof}

We investigate next the affine and elementary involutions. For later
use, we also find their normal forms corresponding to conjugacy by elements in
$C_{\calS}(t)$.

\begin{prop}\label{prop:elem}
In addition to the identity, elementary involutions correspond to the
following classes and normal forms under $C_{\calS}(t)$-conjugacy.
\begin{enumerate}
\item 
	$\left(x,y\right)\rightarrow 
		\left(-x+p(y),y \right),\quad 
		p(y)$ any polynomial.
		
	Normal form: 
		$(x,y)\rightarrow (-x+p(y),y),\quad p(y)= y^{l}+O(y^{l-2})$.
\item 
	$\left(x,y\right)\rightarrow \left(x+p(y),-y+\eta \right),\quad p(y)$
		odd around $\frac{\eta}{2}$.
		
	Normal form: 
		$(x,y)\rightarrow (x+p(y),-y),\quad p(y)$ odd with leading
		coefficient $1$.
\item 
	$\left(x,y\right)\rightarrow 
		\left(-x+p(y),-y+\eta \right),\quad
		p(y)$ even around $\frac{\eta}{2}$.
		
	Normal form:
		$(x,y)\rightarrow (-x+p(y),-y),\quad p(y)$ even with leading
		coefficient $1$.
\end{enumerate}
These normal forms are unique up to replacing $p(y)$ by 
$\zeta\,p(y/\!\zeta)$, where $\zeta$ is any root of unity of order $l-1$
and $l$ is the degree of $p(y)$.
\end{prop}
\begin{proof}
For an elementary automorphism \Eq{eq:elem} we have 
\[
e^{2}\left(x,y\right)=
\left(
\alpha^{2}\,x+\alpha\,p(y) +p(\beta\,y+\eta), 
\beta^{2}\,y+\beta\,\eta +\eta
\right).
\]
Setting $e^{2}=id$ it is easy to see that if $e\ne id$, $e$ has to be of one of
the three classes in the Proposition. Now, defining coordinates $u=ax+b$, 
$v=ay+b$, a simple calculation shows that $e$ can be written in the 
corresponding normal form. Moreover the values of $a$ and $b$ yielding that 
normal form are unique up to $(l-1)\text{-th}$ roots of unity. 
\end{proof}

\begin{rem}It can be proved that every elementary involution is $\E$-conjugate
to one of the affine maps, $(x,y)\rightarrow (\pm x,\pm y)$, where the
coefficients of $x$ and $y$ are conjugacy invariants. However for the purposes 
of this paper we will only normalize the involutions by using conjugacy in
$C_{\calS}(t)$. 
Normal forms for elementary maps are fully discussed in \cite{Friedland89}. 
\end{rem}

\begin{prop}\label{affine}
Let $a$ be an affine, nonelementary automorphism, 
\begin{equation}\label{eq:affine}
a:\left(x,y\right)\rightarrow
\hat{a}\left(x,y\right)+
\left(\xi,\eta\right),
\end{equation}
with $\hat{a}$ linear. Then $a$ is an involution if and only if the eigenvalues
of $\hat{a}$ are $1$ and $-1$ and $\left(\xi, \eta\right)$ is in the
eigenspace of $-1$. 

Furthermore, all affine nonelementary involutions are $\calS$-conjugate to $t$. 
\end{prop}
\begin{proof}
Let be $a$ given by \Eq{eq:affine}. This map is an involution if for every 
$(x,y)$,
\[
	\hat{a}^2\left(x,y\right)+\left(\hat{a}+id\right)
	\left(\xi,\eta\right)=\left(x,y\right).
\]
The above identity holds if and only if $\hat{a}$ is an involution and has $-1$ 
as eigenvalue, with associated eigenvector $\left(\xi,\eta\right)$. 
Taking into account that $\hat{a}$ is not elementary, the condition 
$\hat{a}^2=id$ means that the eigenvalues of $\hat{a}$ must be $1$ and $-1$.
Besides, it can be noted that the eigenspace of $-1$ is generated by 
$(\hat{a}-id)\left(1,0\right)$. This follows from 
$\hat{a}^{2}-id=(\hat{a}+id)(\hat{a}-id)=0$ and the assumption that $a\notin \calS$.

To prove the second part of the Proposition, consider first a linear, 
nonelementary involution $\hat{a}(x,y)$. In that case, taking 
$s(x,y)=x\left(1,0\right)+y\,\hat{a}\left(1,0\right)$, we see that 
$\hat{a}=s\,t\,s^{-1}$.

Next, we show that every affine, non elementary involution \Eq{eq:affine} is
$\calS$-conjugate to its linear part $\hat{a}$. We know that 
$(\xi,\eta)=(\hat{a}-id)(c,0)$ for some scalar $c$. Taking 
$s(x,y)=\left(x+c,y\right)$ it follows that 
$s\,a\,s^{-1}=\hat{a}$ and the proof is complete. 
\end{proof} 

\subsection{Normal forms}
We intend to describe polynomial automorphisms that are reversible by
involutions. Let $g$ be one such automorphism. In that case 
$g=R_{1}\,R_{0}$, where $R_1$ and $R_0$ are involutions. According to 
\Prop{prop:inv}, $R_{i}=g_{i}\,r_{i}\,g_{i}^{-1}$, $i=0,1$, where $r_i$ is an 
elementary or an affine involution and $g_{i}\in \G$. Then $g$ is conjugate to
$r_{1}\,f\,r_{0}\,f^{-1}$ with $f=g_{1}^{-1}\,g_{0}$. If $f\in \calS$, $g$ is
conjugate to the composition of a pair of involutions in $\A\cup \E$. Let us 
consider $f\notin \calS$ so that it can be written as a reduced word,
$f=f_n\,\cdots\,f_1$, $n\geq 1$. However,  
\begin{equation}\label{eq:nored}
	f\,r_{0}\,f^{-1} = 
	f_{n}\,\cdots \,f_{1}\,r_{0}\,f_{1}^{-1}\,\cdots\,f_{n}^{-1},
\end{equation}
is not reduced if $r_0$ and $f_{1}$ are in the same subgroup $\A$ or $\E$. After 
reducing \Eq{eq:nored}, we obtain 
either a map $s_{0}\in \calS$ conjugate to $r_{0}$, or a reduced word
\[
	f_{n}\,\cdots\,f_{k}\,\tilde{r}_{0}\,f_{k}^{-1}\,\cdots \,f_{n}^{-1},
\]
where $\tilde{r}_{0}$ is an affine or an elementary involution, conjugate to 
$r_{0}$. In the last case we see that $g$ is conjugate to 
\[
	f_{k}^{-1}\,\cdots\,f_{n}^{-1}\,r_{1}\,f_{n}\,\cdots\,f_{k}\,\tilde{r}_{0}.
\]
Now, $f_{k}^{-1}\,\cdots\,f_{n}^{-1}\,r_{1}\,f_{n}\,\cdots\,f_{k}$ is not
necessarily a reduced word. This expression reduces either to a map 
$s_{1}\in \calS$ conjugate to $r_{1}$, or to a reduced word, 
$f_{k}^{-1}\,\cdots \,f_{k+l}^{-1}\,\tilde{r}_{1}\,f_{k+l}\,\cdots\,f_{k}$, 
where $\tilde{r}_{1}$ is an affine or an elementary involution, conjugate to
$r_{1}$.
Therefore a polynomial automorphism is reversible if and only if it is 
conjugate to a cyclically-reduced map of one of the following types:
\begin{enumerate}
	\renewcommand{\labelenumi}{\textbf{(R\arabic{enumi}})}
	\item\label{it:trivial}(trivial case)
		$g=\tilde{r}_{1}\,\tilde{r}_{0}\in \A\cup \E$,\ 
		where $\tilde{r}_{0}$, and 
		$\tilde{r}_{1}$ are both affine or both elementary involutions.
	\item\label{it:notrivial}a reduced word, 
		\begin{equation}\label{eq:notrivial}
			g=f_{1}^{-1}\,\cdots\,f_{m}^{-1}
			\,\tilde{r}_{1}\,f_{m}\,\cdots\,f_{1}\,\tilde{r}_{0},
		\end{equation} 
		where $\tilde{r}_{0}, \tilde{r}_{1}$ are involutions. 
		Note that this includes $g$ of the form $\tilde{r}_{1}\,\tilde{r}_{0}$ with 
		$\tilde{r}_0,\tilde{r}_1$ not in the same subgroup $\A$ or $\E$.
\end{enumerate}
An immediate consequence of this structure is that the possible polydegrees for 
conjugacy classes of reversible maps are restricted to be of the form 
$([l_{0}],l_{1},\dots ,l_{m},[l_{m+1}],l_{m},\dots,l_{1})$, the terms in
brackets being optional.

Friedland and Milnor \cite{Friedland89} obtained normal 
forms for conjugacy classes in $\G$ using generalized \Henon 
transformations \Eq{eq:Henon}. Note that \Eq{eq:Henon} is of the form $t\,e$ with 
$e$ an elementary map. If the polynomial $p(y)$ in \Eq{eq:Henon}, has leading
coefficient equal to $1$ and center of mass at $0$, i.e.\ if 
$p(y)=y^{l}+O(y^{l-2})$, we say that the H\'enon transformation is {\em normal}. 
When restricted to the real case we say the transformation is normal if the center
of mass of $p(y)$ is $0$ and the leading coefficient is $\pm 1$. 
Now, according to \cite[Thm. 2.6]{Friedland89}, every cyclically 
reduced map that is not elementary or affine is conjugate to a composition of 
generalized H\'enon transformations. Moreover, with the additional requirement 
that the \Henon transformations be normal, that composition is unique, up to 
finitely many choices. It can be noted that the number of \Henon transformations
in a \Henon normal form equals the semilength of the word. 

For the case of reversible maps, \Henon normal forms do not reflect the 
specific structure of the word. Our next goal is thus to find normal forms 
better adapted to reversible automorphisms. Maps of type \rev{it:trivial} are 
dynamically trivial, thus we study conjugacy classes for maps of type
\rev{it:notrivial}. Our goal is \Th{thm:normal}, which discusses normal forms
for  reversible automorphisms. The following is a preliminary result. 

\begin{lem}\label{lem:henon}
	Given a cyclically-reduced map of the form \Eq{eq:te-form}, there exist 
	diagonal affine automorphisms $s_{m},s_{0}\in C_{\calS}(t)$, such that 
	$s_{m}\,g\,s_{0}^{-1}$ is a composition of normal H\'enon transformations. 
	In other words,
	\[
	s_{m}\,g\,s_{0}^{-1}=t\,\hat{e}_{m}\,\dots\,t\,\hat{e}_{1} \;,
	\]
	where each of the terms $t\,\hat{e}_{i}$ is a normal H\'enon transformation.
	Additionally the terms in the composition are unique up to a finite number of
	choices.
\end{lem}
\begin{proof}
The proof of this result follows closely the methods of \cite{Friedland89}, so
we omit it.
As in their arguments, the normal form is unique up to scaling the polynomials
and the parameters $\delta_{i}$ by $l$-th roots of unity, where
$l=l_{1}\cdots l_{m-1}(l_{m}-1)$ and $l_{i}$ is the degree of $e_{i}$.
\end{proof}

With these results, we can state our main theorem.

\begin{thm}\label{thm:normal}
	Let $g$ be a reversible automorphism of type \rev{it:notrivial}. Then $g$ is 
	conjugate to a cyclically-reduced map of one of the following classes: 
	\begin{list}{}{\setlength{\leftmargin}{1.5cm}\setlength{\labelwidth}{1.25cm}}
	\item[\Naa] a map,
		\begin{equation*}
			(h_{1}^{-1}\,\cdots\,h_{m}^{-1})\,t\,(h_{m}\,\cdots\,h_{1})\,t,
		\end{equation*}
		if $\tilde{r}_{0}$ and $\tilde{r}_{1}$ are both affine;
	\item[\Nea] a map,
		\begin{equation*}
			(h_{1}^{-1}\,\cdots\,h_{m}^{-1})\,e_{m+1}\,(h_{m}\,\cdots\,h_{1})\,t
		\end{equation*}
		if $\tilde{r}_{0}$ and $\tilde{r}_{1}$ belong to different subgroups $\A$ or $\E;$
	\item[\Nee] a map,
		\begin{equation*}
			(t\,h_{1}^{-1}\,\cdots\,h_{m}^{-1})\,e_{m+1}\,
			(h_{m}\,\cdots\,h_{1}\,t)\,e_{0},
		\end{equation*}
		if $\tilde{r}_{0}$ and $\tilde{r}_{1}$ are both elementary;
	\end{list}
	where $h_{i}$ represents a H\'enon transformation and $e_{0},e_{m+1}$ are 
	elementary involutions. 
	Furthermore it can be required that the H\'enon transformations 
	$h_{i}=t\,e_{i},\ i=1,\dots ,m$, as well as the involutions $e_{0},e_{m+1}$ 
	be normalized. In that case the resulting composition is unique up to a finite 
	number of choices.
\end{thm}

We prove this result in \Sec{sec:proof}.

\begin{rem}
	For real polynomial automorphisms, a slight modification of the
	arguments in the proof of \Th{thm:normal} allows us to obtain real 
	normal forms. To do this, we might need to allow the leading coefficient
	of one of the polynomials in the normal H\'enon transformations to be
	$-1$ instead of $1$. This occurs because, as we will see in \Sec{sec:proof},
	the equations that we need to solve to normalize the maps otherwise 
	may not have real solutions.
\end{rem}
The normal forms developed in \Th{thm:normal} give a description of all 
conjugacy classes for reversible polynomial automorphism. They provide also a 
way to verify if a given polynomial automorphism is reversible, by checking if 
it is possible to carry it into any of the normal forms described in the 
theorem. It would be desirable however to obtain a more direct criteria to distinguish
reversible automorphisms. Although that seems difficult in general, we can 
develop conditions for the shorter words. It is clear, for example, that a 
normal \Henon transformation $h=t\,e$ is reversible if and only if $\delta=1,$ 
so that $e$ is a normal involution, or if $\delta=-1$ and $p(y)$ 
is an even
polynomial. These are then the only reversible maps of 
semilength $1$, written as composition of normal \Henon transformations. 
In \Tbl{tbl:short} we summarize the criteria for words of semilength
$m=2,3$ and $4$ when they are given in normal \Henon form, i.e. written as 
composition of normal \Henon transformations, 
$g=h_{m}\,\cdots \,h_{1}$, 
\[
	h_{i}=t\,e_{i}:(x,y)\rightarrow(y,p_{i}(y)-\delta_{i}\,x),\ i=1,\dots,m \;.
\]
The corresponding polydegree is assumed to be $(l_{1},\dots ,l_{m})$. The 
conditions in the table should be understood up to cyclic reorderings of the indexes.
\renewcommand{\arraystretch}{1.25}
\setlength{\tabcolsep}{3mm}
\begin{table}
\begin{center}
\begin{tabular}{|c|c|p{5cm}|p{5cm}|}\hline
	\raisebox{-0.5ex}[0pt]{{\rule[-2mm]{0mm}{7mm}\em Semi-}}&
	\raisebox{-0.5ex}[0pt]{\em Normal Form }\ & & \\
	\raisebox{0.5ex}[0pt]{{\rule[-2mm]{0mm}{7mm}\em length}}&
	\raisebox{0.5ex}[0pt]{\em and Polydegree}&
	\multicolumn{1}{|c|}
	{\raisebox{1.5ex}[0pt]{\em\ Conditions on $\delta_{i}$}}& 
	\multicolumn{1}{|c|}
	{\raisebox{1.5ex}[0pt]
	{\em \ Conditions on $p_{i}$}}\\ \hline\hline
\rule[0mm]{0mm}{5mm}
&\raisebox{-0.25ex}[0pt]{\Naa} & & $p_{2}(y)=c\,\delta_{2}\,p_{1}(c\,y)$,\\
\rule[-2mm]{0mm}{6mm}
\raisebox{1.5ex}[0pt]{2}& $(l_{1},l_{1})$ & 
\raisebox{1.5ex}[0pt]{$\delta_{1}\,\delta_{2}=1$}
& where $c^{l_{1}+1}=\delta_{1}$\\ \hline
& & $\delta_{1}=\delta_{2}=1$ & none \\ \cline{3-4}
& & 
& $p_{1}(y)$ odd \\ \cline{4-4}
2 & \raisebox{1.5ex}[0pt]{\Nee} & 
\raisebox{1.5ex}[0pt]{$\delta_{1}=\delta_{2}=-1$} &
$p_{1}(y),p_{2}(y)$ even \\ \cline{3-4}
& \raisebox{1.5ex}[0pt]{$(l_{1},l_{2})$} & & 
$p_{1}(y)$ even \\ \cline{4-4}
& & \raisebox{1.5ex}[0pt]{$\delta_{1}=1,\ \delta_{2}=-1$} 
& $p_{1}(y)$ odd, $p_{2}(y)$ even \\ \hline\hline
& & $\delta_{1}\,\delta_{2}\,\delta_{3}=1,\ \delta_{3}=\delta_{2}^{l_{1}}$ & 
$p_{3}(y)=\delta_{3}\,p_{1}(\frac{y}{\delta_{2}})$\\ \cline{3-4}
\raisebox{-2.5ex}[0pt]{3}& \raisebox{-1ex}[0pt]{\Nea} & 
\raisebox{-1.5ex}[0pt]
{$\delta_{1}\,\delta_{2}\,\delta_{3}=1,\ \delta_{3}=-(-\delta_{2})^{l_{1}}$} & 
$p_{2}(y)$ odd, \newline 
$p_{3}(y)=-\delta_{3}\,p_{1}(-\frac{y}{\delta_{2}})$\\ 
\cline{3-4}
&\raisebox{1.5ex}[0pt]{$(l_{1},l_{2},l_{1})$} &
\raisebox{-1.5ex}[0pt]
{$\delta_{1}\,\delta_{2}\,\delta_{3}=-1,\ \delta_{3}=-\delta_{2}^{l_{1}}$} &
$p_{2}(y)$ even,\newline
$p_{3}(y)=-\delta_{3}\,p_{1}(\frac{y}{\delta_{2}})$\\
\hline\hline
& \raisebox{-3ex}[0pt]{\Naa} & 
$\delta_{1}\,\delta_{2}\,\delta_{3}\,\delta_{4}=1$, \newline 
$c^{l_{2}+1}=\frac{1}{\delta_{3}}$\ and & 
\raisebox{-2ex}[0pt]{$p_{3}(y)=c\,\delta_{3}\,p_{2}(c\,y)$,} \\ 
\raisebox{1.5ex}[0pt]{4} & \raisebox{-0.5ex}[0pt]{$(l_{1},l_{2},l_{2},l_{1})$} 
& $c^{l_{1}+1}=\delta_{1}^{l_{1}}\,\delta_{4}^{l_{1}+1}$ \newline
for some common $c$.
& $p_{4}(y)=\frac{\delta_{4}}{c}\,p_{1}(\frac{\delta_{1}\delta_{4}}{c}\,y)$\\
\hline
& & $\delta_{1}\,\delta_{3}=1,\ \delta_{2}\,\delta_{4}=1,\ 
\delta_{2}=\delta_{1}^{l_{2}}$ &
$p_{4}(y)=\delta_{4}\,p_{2}(\delta_{1}\,y)$\\ \cline{3-4}
& & $\delta_{1}\,\delta_{3}=1,\ \delta_{2}\,\delta_{4}=1$,\newline
$\delta_{2}=-(-\delta_{1})^{l_{2}}$ &
$p_{1}(y),p_{3}(y)\ \text{odd}$,\newline 
$p_{4}(y)=-\delta_{4}\,p_{2}(-\delta_{1}y)$\\
\cline{3-4}
&\raisebox{-2.5ex}[0pt]{\Nee} & 
$\delta_{1}\,\delta_{3}=1,\ \delta_{2}\,\delta_{4}=1$,\newline 
$\delta_{2}=-\delta_{1}^{l_{2}}$ &
$p_{1}(y),p_{3}(y)\ \text{even}$,\newline 
$p_{4}(y)=-\delta_{4}\,p_{2}(\delta_{1}\,y)$\\
\cline{3-4}
\raisebox{1.5ex}[0pt]{4} & 
\raisebox{-1ex}[0pt]{$(l_{1},l_{2},l_{3},l_{2})$} & 
$\delta_{1}\,\delta_{3}=1,\ \delta_{2}\,\delta_{4}=-1$,\newline
$\delta_{2}=\delta_{1}^{l_{2}}$ &
$p_{3}(y)\ \text{even}$,\newline
$p_{4}(y)=-\delta_{4}\,p_{2}(\delta_{1}\,y)$\\
\cline{3-4}
& & $\delta_{1}\,\delta_{3}=-1,\ \delta_{2}\,\delta_{4}=1$,\newline
$\delta_{2}=-(-\delta_{1})^{l_{2}}$ & 
$p_{1}(y)\ \text{odd},\ p_{3}(y)\ \text{even}$,\newline
$p_{4}(y)=-\delta_{4}\,p_{2}(-\delta_{1}\,y)$\\
\cline{3-4}
& & $\delta_{1}\,\delta_{3}=-1,\ \delta_{2}\,\delta_{4}=-1$,\newline
$\delta_{2}=\delta_{1}^{l_{2}}$ & 
$p_{3}(y)\ \text{odd}$,\newline
$p_{4}(y)=-\delta_{4}\,p_{2}(\delta_{1}\,y)$\\
\hline
\end{tabular}
\caption{\label{tbl:short} 
Conditions for a \Henon normal form map with semilength $m=2,3$, or $4$, to be 
reversible by polynomial involutions.}
\end{center}
\end{table}

\section{Proof of Theorem [\ref{thm:normal}]}\label{sec:proof}
We now proceed to the proof of \Th{thm:normal}.

\begin{proof}
Consider $g$ given by the reduced word \Eq{eq:notrivial}. We can replace all 
affine terms in that expression by $s\,t\,\tilde{s}$, for some 
$s,\tilde{s}\in \calS$. In particular if either $\tilde{r}_{0}$ or $\tilde{r}_{1}$ 
are affine we replace them by $s_{i}\,t\,s_{i}^{-1}$ for some 
$s_{i}\in \calS,\ i=0,1$. This allows us to see that $g$ is conjugate to a word of 
the form
\begin{equation}\label{eq:normal}
	t\,e_{1}^{-1}\,\cdots\,t\,e_{m}^{-1}\,[t\,e_{m+1}]\,
	t\, e_{m}\, \cdots \,t\,e_{1}\,[t\,e_{0}] \;,
\end{equation}
where $e_{0},e_{m+1}$ are elementary involutions. The brackets around
$t\,e_{0}$ and $t\,e_{m+1}$ indicate that those terms may not appear, depending 
on what kind of involutions, affine or elementary, are represented by 
$\tilde{r}_{0}$ and $\tilde{r}_{1}$.
Now, this expression can be written as 
$t\,f^{-1}\,t\,[t\,e_{m+1}]\,f\,[t\,e_{0}]$, 
with $f=t\,e_{m}\,\cdots\,t\,e_{1}$. \Lem{lem:henon} allows us to replace $f$ by
$s_{m}^{-1}\,t\,\hat{e}_{m}\,\cdots\,t\,\hat{e}_{1}\,s_{0}$, with each
$t\,\hat{e}_{i}$ a normal H\'enon transformation and 
$s_{m},s_{0}\in C_{\calS}(t)$. If $f^{-1}$ is similarly replaced, we observe that 
$g$ is conjugate to a map of the
form \Eq{eq:normal}, where the terms $t\,e_{i}$ are now normal \Henon 
transformations, for $i=1,\dots ,m,$ and $e_{0},e_{m+1}$ are any 
elementary involutions. Now, if $\tilde{r}_{0}$ and $\tilde{r}_{1}$ 
are affine, so that $t\,e_{0}$ and $t\,e_{m+1}$ are omitted in \Eq{eq:normal}, 
this proves $g$ is conjugate to the normal form \Naa. However if either of 
these involutions are elementary it is possible to make an additional simplification. 
It should be noted that in such cases the conditions 
$s_{0},s_{m}\in C_{\calS}(t)$ may be unnecessary.

Let us consider the case \Nea, that is when only one of 
$\tilde{r}_{0}, \tilde{r}_{1}$ is an elementary involution. We can assume 
that only $t\,e_{m+1}$ appears in \Eq{eq:normal}, the other case being
equivalent after a cyclic reordering. We can also 
assume that the terms $t\,e_{i},\ i=1,\dots ,m$ are already normal \Henon 
transformations, but $e_{m+1}$ is an arbitrary involution of any of the classes 
described in \Prop{prop:elem}. 
For $i=0,\dots ,m$, we introduce diagonal affine automorphisms 
\begin{equation}\label{eq:diag}
	s_{i}(x_{i},x_{i+1})=(u_{i},u_{i+1}),\quad u_{i}=a_{i}\,x_{i}+b_{i} \;,
\end{equation} 
and replace each term $t\,e_{i},\ i=1,\dots ,m$ by 
$t\,\hat{e}_{i}=s_{i}\,t\,e_{i}\,s_{i-1}^{-1}$.
Now, if on the one hand $s_{0}\in C_{\calS}(t)$, and on the other $t\,e_{i}^{-1}$ 
is replaced by $\phi(s_{i-1})\,t\,e_{i}^{-1}\,\phi(s_{i}^{-1})$ for
$i=1,\dots ,m$, with $\phi$ as given by \Eq{eq:conjugacy}, while 
$t\,e_{m+1}$ becomes replaced by $\phi(s_{m})\,t\,e_{m+1}\,s_{m}^{-1}$, 
we preserve the conjugacy class and the structure of the word. 

For $t\,\hat{e}_{i},\ i=1,\dots ,m,$ to remain normal, it is necessary that 
$b_{i}=0$ and $a_{i+1}=a_{i}^{l_{i}},\ i=1,\dots ,m$. We also need
$a_{0}=a_{1}$ and $b_{0}=b_{1}$ in order to have $s_{0}\in C_{\calS}(t)$. Finally 
the condition that $\hat{e}_{m+1}$ be in normal form yields additional 
equations, 
\[
	\kappa_{m+1}\,a_{m}= a_{m+1}^{l_{m+1}},\qquad 
	\qquad l_{m+1}\,\kappa_{m+1}\,b_{m+1}=\lambda_{m+1}\,a_{m+1} \;,
\]
where 
$p_{m+1}(y)=\kappa_{m+1}\,y^{l_{m+1}}+\lambda_{m+1}\,y^{l_{m+1}-1}+
\text{(lower order terms)}$,
is the polynomial associated to $e_{m+1}$. 
It is not difficult to see that this system of equations gives  
$a_{0}$ up to $l\text{-th}$ roots of unity, for 
$l=l_{1}\dots l_{m-1}(l_{m}l_{m+1}-1)$. All other $a_{i}$ and $b_{i}$ are then 
uniquely determined. 

The remaining case, when both involutions are elementary, follows in a
similar way. 

We have proved existence of normal forms as promised. Uniqueness of those forms
is a consequence of \Th{thm:conjugate}. However some details deserve 
further discussion. Note that to preserve the structure of the word as a 
reversible automorphism, we chose to apply to $t\,e_{i}^{-1}$ the images under
the isomorphism $\phi$ of the maps $s_{i},s_{i-1}$ that modify $t\,e_{i}$. To
discuss uniqueness we have to check if it is possible to apply other diagonal
automorphisms to $t\,e_{i}^{-1}$, and still preserve the structure of the word
as well as the normalizing conditions. 

Suppose that \Eq{eq:normal} is in normal form, and for $i=1,\dots,m$, we
replace $t\,e_{i}$ by $t\,\hat{e}_{i}=s_{i}\,t\,e_{i}\,s_{i-1}^{-1}$, $s_{i}$
given by \Eq{eq:diag}. Consider also diagonal affine automorphisms
$\tilde{s}_{i}(\tilde{x}_{i+1},\tilde{x}_{i})=(\tilde{u}_{i+1},\tilde{u}_{i})$,
with $\tilde{u}_{i}=\tilde{a}_{i}\,\tilde{x}_{i}+\tilde{b}_{i}$, and replace 
$t\,e_{i}^{-1}$ by 
$t\,\tilde{e}_{i}=\tilde{s}_{i-1}\,t\,e_{i}^{-1}\tilde{s}_{i}^{-1},\ 
i=1,\dots,m$. It should be noted that when $t\,e_{0}$ appears it has to be
replaced by $s_{0}\,t\,e_{0}\,\tilde{s}_{0}^{-1}$, otherwise 
$\tilde{s}_{0}$ must be equal to $s_{0};$ similar considerations follow with respect
to $t\,e_{m+1}$. If after these changes the word is still in normal form, with
no need of cyclic reordering, it would be necessary that 
$t\,\tilde{e}_{i}=t\,\hat{e}_{i}^{-1}$ for $i=1,\dots ,m$. This condition is 
equivalent to 
\begin{equation}\label{eq:commute}
	\sigma_{i}\,t\,e_{i}=t\,e_{i}\,\sigma_{i-1} \;, 
\end{equation} 
where $\sigma_{i}=s_{i}^{-1}\,\phi(\tilde{s}_{i})$,
\begin{gather*}
	\sigma_{i}(x,y)= 
	(A_{i}\,x+B_{i},A_{i+1}\,y+B_{i+1}),\quad i=0,\dots m,\\
	A_{i}=\frac{\tilde{a}_{i}}{a_{i}},\quad 
	B_{i}=\frac{\tilde{b}_{i}-b_{i}}{a_{i}},\quad i=0,\dots ,m+1. 
\end{gather*} 
Then \Eq{eq:commute} reduces to 
\begin{equation*}
   \begin{split}
	A_{i+1}&=A_{i-1},\\
	A_{i+1}\,p_{i}(y)+B_{i+1}&=p_{i}(A_{i}\,y+B_{i})-\delta_{i}\,B_{i-1},\
    \end{split}  
\end{equation*} 
for $i=1,\dots ,m.$ It is not difficult to see that the above equations, 
together with the requirement that the terms in the composition be in their 
normal form, imply  $B_{i}=0$ for all $i=0,\dots,m+1$. It also follows that all
$A_{i}=1$ (so that $\sigma_{i}=id$ and $\tilde{s}_{i}=\phi(s_{i})$) unless the
subgroup of roots of unity, 
\begin{equation}\label{eq:roots}
	K=\{\omega \in \C:\, p_{i}(y)=\omega\,p_{i}(\omega\,y),\, i=0,\dots,m+1\}
\end{equation}
is not trivial. Moreover, for normal forms \Nea\ and \Nee\ we also
obtain that all $A_{i}=1$ when the order of $K$ is odd while if the order of $K$
is even either $A_{i}=1$ or $A_{i}=-1$ for every $i$. 

If for the moment we disregard reorderings, the above discussion allows us to
see that the normal forms we have obtained are unique up to the modifications 
we describe next. 
Let us consider again that \Eq{eq:normal} is in normal form and denote by $k$ 
the order of the group $K$ and by $l$ the number 
\begin{equation*}\label{eq:exponent}
	l=l_{0}\dots l_{m-1}(l_{m}l_{m+1}-1),
\end{equation*} 
with $l_{0}$ and $l_{m+1}$ taken equal to $1$ if the corresponding involutions 
do not appear. Let $\zeta$ be any $l\text{-th}$ root of 
unity when $k$ is odd, and any $2l\text{-th}$ root of unity when $k$ even. 
For $i=1,\dots , m+2$ define $a_{i}=\zeta^{l_{0}\dots l_{i-1}}$ and set 
$a_{0}=\zeta^{1-l}$. Then all possible normal forms can be obtained by the 
following modifications: 
\begin{enumerate}
	\item For $i=1,\dots,m+1$ replace the polynomial
	$p_{i}(y)$ related to the elementary map $e_{i}$, by
	$a_{i+1}\,p_{i}(y/\!a_{i})$ and the coefficient
	$\delta_{i}$ by $a_{i+1}\,\delta_{i}/\!a_{i-1}$.
		
	\item If either of the involutions
	$e_{i}(x,y)=(p_{i}(y)-\delta_{i}\,x,\epsilon_{i}\,y),\ i=0$ or
	$i=m+1$ appears, $\epsilon_{i}$ must be replaced by
	$\zeta^{l}\,\epsilon_{i}$.  Besides $p_{0}(y)$ has to be
	replaced by $a_{1}\,p_{0}(y/\!\zeta)$ and the coefficient
	$\delta_{0}$ by $\zeta^{l}\,\delta_{0}$.
\end{enumerate}
It may be noted that for normal form \Naa\ $\zeta$ could be any 
$kl\text{-th}$ root of unity, although in that case
$a_{0}=\zeta^{1-(-1)^{m}l}$. However some further analysis shows that,
depending on the parity of $k$, it suffices to consider $l\text{-th}$ or
$2l\text{-th}$ roots of unity. 

Finally we discuss reorderings. Suppose that $g=g_{2m}\,\dots\,g_{1}$ is a
nontrivial, cyclically-reduced, reversible map.
Then it is possible to factor $g$ as the composition of two involutions, 
\begin{align}
	g&=\left(g_{2m}\,\cdots\,(g_{2k}\,s^{-1})\right)
	\,\left((s\,g_{2k-1})\,\cdots\,g_{1}\right)
	\label{eq:factor}\\
	&=(f_{k+1}^{-1}\,\cdots\,f_{m}^{-1}\,r_{1}\,f_{m}\,\cdots\,f_{k+1})
	(f_{k-1}\,\cdots\,f_{1}\,r_{0}\,f_{1}^{-1}\,\cdots\,f_{k-1}^{-1}),
	\notag
\end{align}
with $s\in \calS$ and $r_{0},r_{1}$ involutions.  It should be noted
that to obtain normal forms we considered a reordering of the terms
that makes the last factor in the reduced word \Eq{eq:factor} of
length $1$.  For any conjugacy class giving rise to normal forms \Naa\
or \Nee, there are two different reorderings of the terms having such
structure, therefore yielding two families of normal forms.  For \Nea\
maps we also required that the last factor corresponds to the affine
involution, therefore in the general case there is only one possible
reordering.  Nevertheless, it is possible that \Eq{eq:factor} can be
factored in more than one way as composition of involutions.  In
other words, the map can have more than two centers of symmetry,
reflecting the existence of different families of reversing
symmetries.  Then other families of normal forms may arise,
corresponding to different choices of symmetry centers.
\end{proof}

\section{Conclusions}\label{sec:concl}
Though every polynomial automorphism is conjugate to a composition of
generalized \Henon maps, we have argued that reversible automorphisms
are more appropriately written in the normal forms given in
\Th{thm:normal}.  There are three possible normal forms depending upon
whether the basic involutory reversors are both elementary \Nee\ both
affine \Naa\ or one of each \Nea.  This is not a complete
classification of reversible automorphisms however, since we have
assumed that the reversors are involutions.  We plan to treat the more
general case in a forthcoming paper \cite{Gomez02}.

There are a number of interesting questions that we have not
investigated.  The automorphisms with polydegree $(l_{1},\dots,l_{n})$
form a manifold of dimension $\sum l_{i} + 6$ \cite{Friedland89}. 
What about the subset of reversible automorphisms?  We have
investigated automorphisms that are reversible in $\G$.  What can one
say about reversible automorphisms that do not have reversors in $\G$? 
Apart from symmetric orbits and their bifurcations, are there other
dynamical properties that distinguish the reversible automorphisms?



\end{document}